\documentstyle[12pt]{article}
\newlength{\extraspace}
\newlength{\extraspaces}
\catcode`\@=11
\def\numberbysection{\@addtoreset{equation}{section}
\def\theequation{\arabic{section}.\arabic{equation}}}
\newcommand{\newsection}[1]{
\vspace{7mm}
\pagebreak[3]
\addtocounter{section}{1}
\setcounter{equation}{0}
\setcounter{subsection}{0}
\setcounter{footnote}{0}
\begin{center}
{\large {\bf \thesection. #1}}
\end{center}
\nopagebreak
\medskip
\nopagebreak
\hspace{3mm}}
\newcommand{\nonu}{\nonumber \\[.5mm]}
\newcommand{\A}{&\!\!\!}

\newcommand{\ket}[1]{\left\vert {#1} \right\rangle}

\newcommand{\be}{\begin{equation}}
\newcommand{\bea}{\begin{eqnarray}}
\newcommand{\eea}{\end{eqnarray}}
\newcommand{\ba}{\begin{array}}
\newcommand{\ea}{\end{array}}
\newcommand{\ee}{\end{equation}}

\newcommand{\br}{\begin{array}}
\newcommand{\er}{\end{array}}
\begin{document}
\addtolength{\baselineskip}{.7mm}
\thispagestyle{empty}
\begin{flushright}
UUITP--18/04 \\
STUPP--04--176 \\
{\tt hep-th/0407106} \\ 
July, 2004
\end{flushright}
\vspace{7mm}
\begin{center}
{\Large{\bf Perturbations and Supersymmetries \\[2mm]
in AdS${}_3$/CFT${}_2$ 
}} \\[20mm] 
{\sc Madoka Nishimura}${}^{\rm a}$\footnote{
\tt e-mail: madoka.nishimura@teorfys.uu.se}
\hspace{1mm} and \hspace{2mm}
{\sc Yoshiaki Tanii}${}^{\rm b}$\footnote{
\tt e-mail: tanii@post.saitama-u.ac.jp} \\[7mm]
${}^{\rm a}${\it Department of Theoretical Physics \\
Uppsala University, Box 803, SE-751 08 Uppsala, Sweden} \\[5mm]
${}^{\rm b}${\it Physics Department, Faculty of Science \\
Saitama University, Saitama 338-8570, Japan} \\[20mm]
{\bf Abstract}\\[7mm]
{\parbox{13cm}{\hspace{5mm}
Symmetry breaking by perturbations in the AdS/CFT 
correspondence is discussed. 
Perturbations of vector fields to the 
AdS${}_3$ $\times$ S${}^3$ solution of the six-dimensional 
${\cal N}=(4,4)$ supergravity are considered. 
These perturbations are identified as descendents of chiral 
primary operators of a two-dimensional ${\cal N}=(4,4)$ 
CFT with conformal weight $(2,2)$ or $(1,1)$. 
We examine unbroken symmetries by the perturbations 
in the CFT side as well as in the supergravity side and 
find the same result: 
the ${\cal N}=(4,2)$ or ${\cal N}=(2,4)$ Poincar\'e supersymmetry 
for the $(2,2)$ perturbation and 
the ${\cal N}=(0,4)$ or ${\cal N}=(4,0)$ superconformal symmetry 
for the $(1,1)$ perturbation. 
}}
\end{center}
\vfill
\newpage
\setcounter{section}{0}
\setcounter{equation}{0}
\numberbysection
%
\newsection{Introduction}
%
In the original context of the AdS/CFT correspondence 
\cite{MAL,GKP,WITTEN} string theories or supergravities 
in the AdS space describe field theories on the boundary 
with the conformal symmetry and a large extended supersymmetry. 
(For a review, see ref.\ \cite{AGMOO}.) 
To apply it to more realistic models one has to consider 
theories with lower supersymmetries. 
One should understand supersymmetry breaking in both of the 
supergravity side and the field theory side. 
\par
One of the approaches to obtain the AdS/CFT correspondence 
for lower supersymmetric cases is to modify AdS solutions of 
supergravities by adding a perturbation. In ref.\ \cite{PS}, 
for instance,  a perturbation of the three-form flux was added 
to the AdS${}_5$ $\times$ S${}^5$ solution, which breaks 
${\cal N}=4$ supersymmetry to ${\cal N}=1$. The perturbation is 
a solution of the linearized field equation around the 
AdS${}_5$ $\times$ S${}^5$ background. This perturbation 
corresponds to fermion mass terms of the three ${\cal N}=1$ 
chiral multiplets in the ${\cal N}=4$ super Yang-Mills theory 
and polarizes D3-branes into 5-branes \cite{MYERS,TVR}. 
It is easy to see how these mass terms break the supersymmetry 
in the field theory side. 
Furthermore, supersymmetry breaking by the perturbation 
was also studied in the supergravity side \cite{GP,NT4} 
by examining supertransformations of the fermionic fields. 
The results of supersymmetry breaking are consistent 
in the field theory side and in the supergravity side. 
\par
A similar supersymmetry breaking by a perturbation was 
discussed for a two-dimensional CFT 
and its dual supergravity solution. 
In ref.\ \cite{NISHI} solutions of the linearized field 
equations of vector fields around the AdS${}_3$ $\times$ 
S${}^3$ solution of the six-dimensional ${\cal N}=(4,4)$ 
supergravity were obtained. 
This six-dimensional supergravity is an effective theory 
of the type IIB superstring compactified on T${}^4$ 
with the size of T${}^4$ much smaller than those of 
AdS${}_3$ and S${}^3$. 
By adding these solutions of vector fields as a perturbation 
the ${\cal N}=(4,4)$ superconformal symmetry of 
the two-dimensional dual field theory is broken. 
A preliminary analysis in the supergravity side showed that 
there are cases in which it is broken to ${\cal N}=(4,0)$. 
In contrast to the above AdS${}_5$ $\times$ S${}^5$ case 
the physical meaning of the perturbations in the field theory 
side is not clear in this case. 
By this reason supersymmetry breaking in the field theory side 
was not studied in ref.\ \cite{NISHI}. 
\par
The purpose of the present paper is to study supersymmetry 
breaking by the perturbations in this model in more detail. 
We first identify operators of the two-dimensional CFT 
corresponding to the perturbations of the supergravity solution. 
The relation between operators of the CFT and linearized 
solutions of the supergravity was studied in 
refs.\ \cite{MS,DKSS,BOER,APT}. We use these results to find that 
the perturbations correspond to certain descendents of chiral primary 
operators of the ${\cal N}=(4,4)$ superconformal field theory, 
which have conformal weight $(h,\bar{h})=(2,2)$ or $(1,1)$. 
We then examine breaking of supersymmetry as as well as of 
bosonic symmetries by perturbations of these operators 
in the CFT side. 
We find that the unbroken symmetry is 
the ${\cal N}=(4,2)$ or ${\cal N}=(2,4)$ Poincar\'e supersymmetry 
for the $(2,2)$ perturbation, while it is 
the ${\cal N}=(0,4)$ or ${\cal N}=(4,0)$ superconformal symmetry 
for the $(1,1)$ perturbation. 
Finally, we examine symmetry breaking in the supergravity side 
by studying supertransformations of the fermionic fields. 
The unbroken symmetries are in complete agreement with those 
in the CFT side. 
The result in this paper may be regarded as another non-trivial 
evidence in support of the AdS/CFT correspondence. 
\par
In the next section we review the AdS${}_3$ $\times$ S${}^3$ 
solution of the six-dimensional ${\cal N}=(4,4)$ supergravity and 
its symmetries. In sect.\ 3 we give the perturbations around this 
solution obtained in ref.\ \cite{NISHI}. 
In sect.\ 4 we first identify operators in the CFT corresponding 
to these perturbations. Then, we examine unbroken symmetries by 
the perturbations in the CFT side. 
In sect.\ 5 we examine unbroken symmetries in the supergravity side 
and show that they precisely coincide with those in the CFT side. 
In Appendix we give our conventions of SO(4) and SO(5) gamma matrices 
used in the text. 
\par
%
\newsection{AdS${}_3$ $\times$ S${}^3$ background}
%
We first recall the AdS${}_3$ $\times$ S${}^3$ solution 
of the six-dimensional supergravity. 
The six-dimensional ${\cal N}=(4,4)$ supergravity 
\cite{CREMMER,TANII} has a rigid SO(5,5) 
symmetry and a local SO(5) $\times$ SO(5) symmetry. 
The field content of the theory is a vielbein $e_M{}^{\hat{M}}$, 
five antisymmetric tensor fields $B^m_{MN}$, 16 vector fields 
$A_M^{\tilde{\mu}\dot{\tilde{\mu}}}$, 25 scalar fields 
$\phi_{\tilde{\mu}\dot{\tilde{\mu}}}^{\alpha\dot{\alpha}}$, 
eight Rarita-Schwinger fields $\psi_{+M\alpha}$, 
$\psi_{-M\dot{\alpha}}$ and 40 spinor fields 
$\chi_{+a\dot{\alpha}}$, $\chi_{-\dot{a}\alpha}$. 
The indices $M, N, \cdots$ and $\hat{M}, \hat{N}, \cdots$ are 
six-dimensional world and local Lorentz indices, respectively. 
Other indices take values $m, a, \dot{a} = 1, \cdots, 5$ and 
$\tilde\mu, \dot{\tilde\mu}, \alpha, \dot\alpha = 1, \cdots, 4$. 
A pair of indices $\tilde\mu\dot{\tilde\mu}$ represent a spinor 
index of SO(5,5), while $a, \dot{a}$ and $\alpha, \dot\alpha$ 
represent vector and spinor indices of SO(5) $\times$ SO(5), 
respectively. 
The field strengths of the antisymmetric tensor fields 
and their duals belong to {\bf 10} of SO(5,5). 
The fermionic fields are SO(5)-symplectic 
Majorana-Weyl spinors and the signs on the fields denote the 
chiralities. The scalar fields take values on the coset space 
SO(5,5)/(SO(5) $\times$ SO(5)). 
\par
We are only interested in the fields $e_M{}^{\hat{M}}$, $B_{MN}^m$ 
and $A_M^{\tilde\mu\dot{\tilde\mu}}$ and set other fields to zero 
except $\phi_{\tilde{\mu}\dot{\tilde{\mu}}}^{\alpha\dot{\alpha}} 
= \delta^\alpha_{\tilde\mu} \delta^{\dot\alpha}_{\dot{\tilde\mu}}$. 
By this scalar field background the rigid SO(5,5) and 
local SO(5) $\times$ SO(5) symmetries are broken to a rigid 
SO(5) $\times$ SO(5) symmetry, and the indices $\tilde\mu$ 
and $\alpha$, $\dot{\tilde\mu}$ and $\dot\alpha$ are identified, 
respectively. The local supertransformation of the fermionic 
fields \cite{TANII}\footnote{There are several misprints in 
ref.\ \cite{TANII}. 
The left-hand sides of eq.\ (17) should be $\tilde{H}_{1\pm MNP}^m$. 
The right-hand side of the first equation of eq.\ (24) should be 
$F^m - \tilde{H}_{1+}^m - \tilde{H}_{1-}^m$. 
There should be a minus sign on the right-hand side of the third 
equation of eq.\ (24). 
The coefficient of the fifth term of eq.\ (21) should not be 
$-{1 \over 4}i$ but $-{1 \over 2}i$. 
The eighth term of eq.\ (21) should be 
$-{1 \over 3} ( F_+^a \cdot H_{0+}^a 
+ F_-^{\dot{a}} \cdot H_{0-}^{\dot{a}} )$.} 
in this background becomes 
\begin{eqnarray}
\delta \psi_{+M\alpha} 
\A = \A D_M \epsilon_{+\alpha} 
+ {1 \over 4} H^a_{+MNP} (\gamma_a)_\alpha{}^\beta 
\Gamma^{NP} \epsilon_{+\beta}
+ {3 \over 4} G_{MN\alpha\dot{\beta}} \Gamma^N 
\epsilon_-^{\dot{\beta}}
- {1 \over 8} G_{NP\alpha\dot{\beta}} \Gamma_M{}^{NP} 
\epsilon_-^{\dot{\beta}}, \nonu
\delta \psi_{-M\dot{\alpha}} 
\A = \A D_M \epsilon_{-\dot{\alpha}} 
- {1 \over 4} H^{\dot{a}}_{-MNP} 
(\gamma_{\dot{a}})_{\dot{\alpha}}{}^{\dot{\beta}}
\Gamma^{NP} \epsilon_{-\dot{\beta}}
+ {3 \over 4} G_{MN\beta\dot{\alpha}} \Gamma^N \epsilon_+^\beta
- {1 \over 8} G_{NP\beta\dot{\alpha}} \Gamma_M{}^{NP} 
\epsilon_+^\beta, \nonu
\delta \chi_{+a\dot{\alpha}} 
\A = \A - {1 \over 12} H_{+aMNP} \Gamma^{MNP} 
\epsilon_{-\dot{\alpha}}
- {1 \over 4} G_{MN\beta\dot{\alpha}} \Gamma^{MN} 
\epsilon_+^\alpha (\gamma_a)_\alpha{}^\beta, \nonu
\delta \chi_{-\dot{a}\alpha} 
\A = \A {1 \over 12} H_{-\dot{a}MNP} \Gamma^{MNP} 
\epsilon_{+\alpha}
- {1 \over 4} G_{MN\alpha\dot{\beta}} \Gamma^{MN} 
\epsilon_-^{\dot{\alpha}} 
(\gamma_{\dot{a}})_{\dot{\alpha}}{}^{\dot{\beta}}, 
\label{susytransformation}
\end{eqnarray}
where the transformation parameter $\epsilon_{+\alpha}$ and 
$\epsilon_{-\dot{\alpha}}$ are SO(5)-symplectic Majorana-Weyl 
spinors. 
%
$\Gamma^{\hat{M}}$ are gamma matrices of the six-dimensional 
Lorentz group SO(1,5), while $\gamma^a$, $\gamma^{\dot{a}}$ are 
those of SO(5) $\times$ SO(5). 
The field strengths of the tensor and vector fields are defined as 
\begin{eqnarray}
H_{MNP}^m \A = \A 3 \partial_{[M} B_{NP]}^m
+ {3 \over 2} G_{[MN}^{\alpha\dot{\alpha}} 
A_{P]\beta\dot{\alpha}} (\gamma^m)_\alpha{}^\beta 
- {3 \over 2} G_{[MN}^{\alpha\dot{\alpha}} A_{P]\alpha\dot{\beta}} 
(\gamma^m)_{\dot{\alpha}}{}^{\dot{\beta}}, \nonu
G_{MN}^{\alpha\dot{\alpha}} 
\A = \A 2 \partial_{[M} A_{N]}^{\alpha\dot{\alpha}}. 
\label{fieldstrength}
\end{eqnarray}
$H_+^a$ and $H_-^{\dot{a}}$ are self-dual and anti self-dual 
part of $H^m$ with $m=a$ or $m=\dot{a}$, and transform as 
$({\bf 5}, {\bf 1})$ and $({\bf 1}, {\bf 5})$ under the rigid 
SO(5) $\times$ SO(5) respectively. 
$G_{MN}^{\alpha\dot{\alpha}}$ satisfies a doubly-symplectic 
reality condition 
\begin{equation}
(G_{MN}^{\alpha\dot{\alpha}})^* 
= (\Omega^{-1})_{\alpha\beta} (\Omega^{-1})_{\dot{\alpha}\dot{\beta}}
G_{MN}^{\beta\dot{\beta}}, 
\label{reality}
\end{equation}
where $\Omega^{\alpha\beta}$ and $\Omega^{\dot{\alpha}\dot{\beta}}$
are antisymmetric SO(5) charge conjugation matrices (See Appendix.). 
\par
The AdS${}_3$ $\times$ S${}^3$ solution has a metric 
\begin{eqnarray}
ds^2 \A = \A Z(r)^{-1} dx^\mu dx^\nu \eta_{\mu\nu} 
+ Z(r) dx^i dx^j \delta_{ij} \nonu
\A = \A Z(r)^{-1} dx^\mu dx^\nu \eta_{\mu\nu} 
+ Z(r) dr^2 + R^2 d\Omega_3^2 
\label{solofmetric}
\end{eqnarray}
and a self-dual field strength with non-vanishing components 
\begin{equation}
H_{\mu\nu i}^a = R^{-2} {\cal S}^a \epsilon_{\mu\nu} x^i,  \qquad
H_{ijk}^a = - r^{-4} R^2 {\cal S}^a \epsilon_{ijkl} x^l,  
\label{solof3form}
\end{equation}
where $r^2=x^ix^i$, $Z(r) = {R^2 \over r^2}$, and 
$d\Omega_3^2$ is the metric of S${}^3$ of unit radius. 
We have split the six-dimensional world index as 
$M=(\mu, i)$ ($\mu=0,1$; $i=2,3,4,5$). 
The antisymmetric $\epsilon_{\mu\nu}$ and $\epsilon_{ijkl}$ 
are chosen as $\epsilon_{01}=+1=\epsilon_{2345}$. 
${\cal S}^a$ is a constant vector of unit length 
${\cal S}^a {\cal S}^a = 1$. We choose ${\cal S}^5=1$ and 
${\cal S}^a=0$ ($a=1,2,3,4$) without losing generality. 
The constant parameter $R$ denotes the radius of AdS${}_3$ 
and S${}^3$. The metric (\ref{solofmetric}) gives a vielbein 
and a spin connection as 
\begin{eqnarray}
e_\mu{}^{\hat{\mu}} 
\A = \A \delta_\mu^{\hat{\mu}} Z^{-{1 \over 2}}, \qquad
e_i{}^{\hat{i}} = \delta_i^{\hat{i}} Z^{1 \over 2}, \nonu
\omega_\mu{}^{\hat{\nu}\hat{i}} 
\A = \A {x^i \over R^2} \delta_\mu^{\hat{\nu}} 
\delta_i^{\hat{i}}, \qquad
\omega_i{}^{\hat{j}\hat{k}} 
= - {x^l \over r^2} ( 
\delta_i^{\hat{j}} \delta_l^{\hat{k}} 
- \delta_i^{\hat{k}} \delta_l^{\hat{j}} ), 
\label{eomega}
\end{eqnarray}
where $\hat{\mu}, \hat{\nu}, \cdots$ and 
$\hat{i}, \hat{j}, \cdots$ denote local Lorentz indices. 
It is convenient to decompose the six-dimensional gamma 
matrices $\Gamma^{\hat{M}}$ as 
\begin{eqnarray}
\Gamma^{\hat{\mu}} 
& = & \hat{\gamma}^{\hat{\mu}} \otimes \bar{\gamma}_{\rm{4D}}, 
\nonu
\Gamma^{\hat{i}} & = & 1 \otimes \hat{\gamma}^{\hat{i}}, 
\end{eqnarray}
where $\hat\gamma^{\hat{\mu}}$ and $\hat\gamma^{\hat{i}}$ 
are gamma matrices of SO(1,1) and SO(4) respectively, and 
we have defined 
\begin{equation}
\bar{\gamma}_{\rm{2D}} 
= \hat{\gamma}^{\hat{0}} \hat{\gamma}^{\hat{1}}, 
\qquad
\bar{\gamma}_{\rm{4D}} 
= \hat{\gamma}^{\hat{2}} \hat{\gamma}^{\hat{3}} 
\hat{\gamma}^{\hat{4}} \hat{\gamma}^{\hat{5}}.
\label{chiralitymatrix}
\end{equation}
We use the explicit representations of the SO(4) gamma 
matrices given in Appendix. 
\par
The AdS${}_3$ $\times$ S${}^3$ solution (\ref{solofmetric}), 
(\ref{solof3form}) has bosonic and fermionic symmetries. 
The rigid SO(5) $\times$ SO(5) symmetry is broken to a rigid 
SO(4) $\times$ SO(5) by the non-vanishing value of 
$H_+^a \propto {\cal S}^a$ in $({\bf 5}, {\bf 1})$. 
The first factor SO(4) $\sim$ SU(2) $\times$ SU(2) 
corresponds to the automorphism group of the ${\cal N}=(4,4)$ 
superconformal algebra to be discussed in sect.\ 4,  
while the second factor SO(5) will not play important role 
in the following discussion. 
The solution is also invariant under the isometry of 
AdS${}_3$ $\times$ S${}^3$. 
The isometry of S${}^3$ is SO(4) $\sim$ SU(2) $\times$ SU(2), 
which acts on $x^i$ as SO(4) rotations. 
This symmetry corresponds to the SU(2) $\times$ SU(2) generators 
$J_0^I$, $\tilde{J}_0^{I'}$ in the ${\cal N}=(4,4)$ superconformal 
algebra in sect.\ 4. 
The isometry of AdS${}_3$ is SO(2,2) $\sim$ SO(2,1) $\times$ 
SO(2,1). It is generated by Killing vectors $\xi^M$ 
\begin{equation}
\xi^\mu = \zeta^\mu -{1 \over 4}{R^4 \over r^2} 
\eta^{\mu\nu}\partial_\nu \partial_\rho \zeta^\rho, \qquad
\xi^i = - {1 \over 2} x^i \partial_\rho \zeta^\rho,
\label{killing}
\end{equation}
where $\zeta^\mu(x^\nu)$ is an arbitrary vector satisfying 
\begin{equation}
\partial_\mu \zeta^\rho \eta_{\rho\nu} 
+ \partial_\nu \zeta^\rho \eta_{\rho\mu} 
= \partial_\rho \zeta^\rho \eta_{\mu\nu}, \qquad
\partial_\mu \partial_\nu \partial_\rho \zeta^\rho = 0. 
\end{equation}
Thus, $\zeta^\mu$ is a two-dimensional conformal Killing vector 
which is quadratic in $x^\mu$. 
This symmetry corresponds to Virasoro generators 
$L_m$, $\tilde{L}_m$ ($m=\pm1, 0$) in the 
superconformal algebra. 
\par
Finally, supersymmetries preserved by this solution are 
given by the parameters $\epsilon_- = 0$ and $\epsilon_+$ 
satisfying 
\begin{equation}
D_M \epsilon_{+\alpha} 
+ {1 \over 4} H^a_{+MNP} (\gamma_a)_\alpha{}^\beta 
\Gamma^{NP} \epsilon_{+\beta} = 0. 
\label{killingspinor}
\end{equation}
This condition comes from the vanishing of the 
supertransformations of the fermionic fields 
(\ref{susytransformation}). Substituting 
eqs.\ (\ref{solof3form}), (\ref{eomega}) into 
eq. (\ref{killingspinor}) it becomes 
\begin{eqnarray}
\left[ \partial_\mu - {1 \over 2R} \hat{\gamma}_\mu 
\hat{\gamma}^{\hat{r}} \bar\gamma_{4D} 
( 1 - \bar\gamma_{2D} \gamma_5 ) \right]
\epsilon_+ \A = \A 0, \nonu
\left[ \partial_i - {x^i \over 2r^2} \bar{\gamma}_{2D} \gamma_5 
- {x^j \over 2r^2} \hat{\gamma}_{\hat{i}\hat{j}} 
(1- \bar{\gamma}_{2D} \gamma_5) \right] \epsilon_+ \A = \A 0, 
\label{killingspinor2}
\end{eqnarray}
where $\hat{\gamma}^{\hat{r}} = {x^i \over r}\hat{\gamma}^{\hat{i}}$, 
and $\gamma_5$ is the fifth matrix of the SO(5) gamma matrices 
$\gamma_a$. 
The general $\epsilon_+$ satisfying these equations is a sum of 
\begin{eqnarray}
\epsilon_+^{(++)} 
\A = \A r^{1 \over 2} \eta^{(++)}, \nonu
\epsilon_+^{(-+)} 
\A = \A - {1 \over 2} r^{1 \over 2} \hat{\gamma}^{\hat{r}} 
\hat{\gamma}^\mu \partial_\mu \eta^{(++)}, \nonu
\epsilon_+^{(--)} 
\A = \A r^{1 \over 2} \eta^{(--)}, \nonu
\epsilon_+^{(+-)} 
\A = \A {1 \over 2} r^{1 \over 2} \hat{\gamma}^{\hat{r}} 
\hat{\gamma}^\mu \partial_\mu \eta^{(--)}, 
\label{susyparameter}
\end{eqnarray}
where the suffix $(\pm\pm)$ on $\epsilon_+$ and $\eta$ denotes 
eigenvalues of $\bar{\gamma}_{4D}$ and $\gamma_5$. 
$\eta^{(\pm\pm)}$ are two-dimensional conformal Killing spinors 
\begin{eqnarray}
\eta^{(++)}(x^+)
\A = \A \epsilon_0^{(++)}
+ {\sqrt{2} \over R} \epsilon_1^{(++)} x^+, \nonu
\eta^{(--)}(x^-)
\A = \A \epsilon_0^{(--)}
+ {\sqrt{2} \over R} \epsilon_1^{(--)} x^-, 
\label{susyparameter2}
\end{eqnarray}
where $x^\pm = {1 \over \sqrt{2}} (x^0\pm x^1)$, and 
$\epsilon_0^{(\pm\pm)}$ and $\epsilon_1^{(\pm\pm)}$ 
are arbitrary constant spinors with given $\bar{\gamma}_{4D}$ 
and $\gamma_5$ eigenvalues. 
Note that $\eta^{(\pm\pm)}$ must be linear in $x^\pm$. 
\par
The boundary of AdS${}_3$ at infinity, on which the CFT is defined, 
is a cylinder. We will use the coordinates of the cylinder 
when we discuss the CFT in sect.\ 4. 
To compare the supergravity side and the CFT side 
we need a relation between the coordinates $x^\mu$ 
in eq.\ (\ref{solofmetric}) and the coordinates of the cylinder 
$\tau$, $\sigma$ 
($-\infty < \tau < \infty$, $0 \leq \sigma \leq 2\pi$), 
which is given by (see, e.g. ref.\ \cite{AGMOO}) 
\begin{equation}
e^{i(\tau\pm\sigma)}
= {R + i (x^0 \pm x^1) \over R - i (x^0 \pm x^1)}.
\end{equation}
Going to the Euclidean signature this relation becomes 
\begin{equation}
z = {1+w \over 1-w}, 
\label{zw}
\end{equation}
where 
\begin{equation}
z = e^{\tau_E + i \sigma}, \qquad
w = {1 \over R} (t_E + i x^1), 
\label{zw2}
\end{equation}
and $\tau_E = i \tau$, $t_E = i x^0$ are Euclidean time coordinates. 
In terms of the coordinate $z$ the conformal Killing 
spinors (\ref{susyparameter2}) become 
\begin{eqnarray}
\eta^{(++)}(z)
\A = \A {1 \over \sqrt{2}} (1+z) \epsilon_0^{(++)} 
+ {1 \over \sqrt{2}} i (1-z) \epsilon_1^{(++)}, \nonu
\eta^{(--)}(\bar{z})
\A = \A {1 \over \sqrt{2}} (1+\bar{z}) \epsilon_0^{(--)} 
+ {1 \over \sqrt{2}} i (1-\bar{z}) \epsilon_1^{(--)}, 
\label{susyparameter3}
\end{eqnarray}
where we have used the fact that $\eta^{(\pm\pm)}$ transforms as 
a primary field of weight $-{1 \over 2}$ under the conformal
transformation (\ref{zw}) in the same way as for superconformal 
ghosts in the superstring \cite{FMS}. 
{}From the expression (\ref{susyparameter3}) we see that 
the parameters $\epsilon_0^{(++)}$ and $\epsilon_1^{(++)}$ 
correspond to combinations of the supercharges 
$G_{-{1 \over 2}} + G_{1 \over 2}$ and 
$G_{-{1 \over 2}} - G_{1 \over 2}$ 
in the superconformal algebra in sect.\ 4 respectively. 
Similarly, $\epsilon_0^{(--)}$ and $\epsilon_1^{(--)}$ 
correspond to 
$\tilde{G}_{-{1 \over 2}} + \tilde{G}_{1 \over 2}$ and 
$\tilde{G}_{-{1 \over 2}} - \tilde{G}_{1 \over 2}$ respectively. 
\par
%
\newsection{Perturbations in supergravity}
%
In ref.\ \cite{NISHI} perturbations of the vector fields 
were obtained which satisfy the linearized field equations 
around the AdS${}_3$ $\times$ S${}^3$ solution 
(\ref{solofmetric}), (\ref{solof3form}). 
We consider the linear order in these perturbations. 
There is no back reaction to other fields to this order.
\par
To show the perturbations we introduce self-dual and anti 
self-dual two-forms $T_2 = {1 \over 2} T_{ij} dx^i \wedge dx^j$ 
satisfying 
\begin{equation}
*_4 T_2 = \pm T_2, 
\end{equation}
where $*_4$ is the Hodge dual for the flat metric $\delta_{ij}$. 
The general forms of these two-forms are 
\begin{equation}
T_2 = m_1 dz^1 \wedge d\bar{z}^2 + m_2 d\bar{z}^1 \wedge dz^2 
+ m_3 ( dz^1 \wedge d\bar{z}^1 - dz^2 \wedge d\bar{z}^2 )   
\label{sdt}
\end{equation}
for the self-dual case and 
\begin{equation}
T_2 = m_1 dz^1 \wedge dz^2 + m_2 d\bar{z}^1 \wedge d\bar{z}^2 
+ m_3 ( dz^1 \wedge d\bar{z}^1 + dz^2 \wedge d\bar{z}^2 ). 
\label{asdt}
\end{equation}
for the anti self-dual case, where $m_1$, $m_2$, $m_3$ are constant 
coefficients and we have introduced 
the complex coordinates 
\begin{equation}
z^1={1 \over \sqrt{2}} (x^2+ix^4), \qquad 
z^2={1 \over \sqrt{2}} (x^3+ix^5). 
\end{equation}
Under the isometry SU(2) $\times$ SU(2) of S${}^3$ these 
two-forms transform 
as ({\bf 3}, {\bf 1}) and ({\bf 1}, {\bf 3}) respectively. 
In particular, the $m_1$ terms in eqs.\ (\ref{sdt}), (\ref{asdt}) 
represent the highest weight state of each SU(2). 
We also define a two-form $V_2$ from $T_2$ with components 
\begin{equation}
V_{ij} = {x^k \over r^2} (x^iT_{kj} + x^jT_{ik}). 
\end{equation}
\par
The perturbations satisfying the field equations were given 
in terms of $T_2$ and $V_2$ in ref.\ \cite{NISHI}. 
We give the field strengths  $G_2$ of the vector fields 
in eq.\ (\ref{fieldstrength}). 
For the self-dual case $*_4 T_2 = T_2$ there are two pairs 
of solutions 
\begin{eqnarray}
G^{(+)}_2 \A = \A {1 \over 2} T_2 (1-\gamma_5), \nonu 
G^{(-)}_2 \A = \A {1 \over 2} r^{-6} (T_2 - 3V_2) (1-\gamma_5) 
\label{sd1}
\end{eqnarray}
and 
\begin{eqnarray}
G^{(+)}_2 \A = \A {1 \over 2} r^{-2} (T_2 -V_2) (1+\gamma_5), \nonu
G^{(-)}_2 \A = \A {1 \over 2} r^{-4} (T_2 - 2V_2) (1+\gamma_5). 
\label{sd2}
\end{eqnarray}
Similarly, for the anti self-dual case $*_4 T_2 = - T_2$ 
there are two pairs of solutions 
\begin{eqnarray}
G^{(+)}_2 \A = \A {1 \over 2} T_2 (1+\gamma_5), \nonu
G^{(-)}_2 \A = \A {1 \over 2} r^{-6} (T_2 - 3V_2) (1+\gamma_5) 
\label{asd1}
\end{eqnarray}
and 
\begin{eqnarray}
G^{(+)}_2 \A = \A {1 \over 2} r^{-2} (T_2 - V_2) (1-\gamma_5), \nonu
G^{(-)}_2 \A = \A {1 \over 2} r^{-4} (T_2 - 2V_2) (1-\gamma_5). 
\label{asd2}
\end{eqnarray}
By the reality condition of $G_{ij}$ (\ref{reality}) the 
coefficients of $T_2$ must satisfy 
\begin{equation}
(m_1^{\alpha\dot{\alpha}})^* 
= (\Omega^{-1})_{\alpha\beta} 
(\Omega^{-1})_{\dot{\alpha}\dot{\beta}} \, 
m_2^{\beta\dot{\beta}}, \qquad
(m_3^{\alpha\dot{\alpha}})^* 
= - (\Omega^{-1})_{\alpha\beta} 
(\Omega^{-1})_{\dot{\alpha}\dot{\beta}} \, 
m_3^{\beta\dot{\beta}}. 
\end{equation}
For each pair $G^{(+)}_2$ represents a perturbation in the CFT 
by a operator, while $G^{(-)}_2$ represents the vacuum 
expectation value of the operator \cite{BKL,BKLT}. 
We will examine symmetries preserved by $G^{(+)}_2$. 
\par
%
\newsection{Perturbations in CFT}
%
The AdS${}_3$ $\times$ S${}^3$ $\times$ T${}_4$ supergravity 
background corresponds to a two-dimensional ${\cal N}=(4,4)$ 
superconformal field theory \cite{MAL}. 
This CFT is described as a deformation of the 
supersymmetric sigma model with a target space $T_4{}^N/S_N$ 
\cite{VAFA,SV}. For details of this theory see, e.g. 
ref.\ \cite{DMW}. The perturbations of the supergravity solution 
discussed in the previous section correspond to certain operators 
in the CFT. In this section we identify these operators and 
examine unbroken symmetries by these operators. 
We do not need the detailed properties of the operators but 
the fact that they are operators corresponding to descendents 
of chiral primary states. 
\par
${\cal N}=(4,4)$ superconformal field theories have two copies 
of the ${\cal N}=4$ super Virasoro algebra for the holomorphic 
and the anti-holomorphic parts. The ${\cal N}=4$ super Virasoro 
algebra for the holomorphic part consists of Virasoro generators 
$L_n$, SU(2) currents $J_n^I$ ($I=1,2,3$) and supercharges 
$G_r^{A\dot{A}}$ ($A=1,2$; $\dot{A}=\dot{1}, \dot{2}$). 
The mode indices take values 
$n \in {\bf Z}$ and $r \in {\bf Z}+{1 \over 2}$ corresponding 
to the anti-periodic boundary condition on fermionic fields. 
The (anti-)commutation relations of this algebra are 
\begin{eqnarray}
[ L_m, L_n ] 
\A = \A (m-n) L_{m+n} 
+ {c \over 12} m (m^2-1) \delta_{m+n,0}, \nonu
\{ G_r^{A\dot{A}}, G_s^{B\dot{B}} \} 
\A = \A \epsilon^{\dot{A}\dot{B}} \epsilon^{AB} L_{r+s} 
+ (r-s) \epsilon^{\dot{A}\dot{B}} (\sigma^I)^{AB} J_{r+s}^I 
+ {c \over 6} \left( r^2 - {1 \over 4} \right) 
\epsilon^{\dot{A}\dot{B}} \epsilon^{AB} \delta_{r+s,0}, \nonu
[ J_m^I, J_n^J ] 
\A = \A i \epsilon^{IJK} J_{m+n}^K 
+ {c \over 12} m \delta^{IJ} \delta_{m+n, 0}, \nonu
[ L_m, G_r^{A\dot{A}} ] 
\A = \A \left( {1 \over 2}m - r \right) 
G_{m+r}^{A\dot{A}}, \nonu
[ L_m, J_n^I ] 
\A = \A - n J_{m+n}^I, \nonu
[ J_m^I, G_r^{A\dot{A}} ] 
\A = \A - {1 \over 2} (\sigma^I)^A{}_B 
G_{m+r}^{B\dot{A}}, 
\label{n4algebra}
\end{eqnarray}
where $\epsilon^{AB}$ and $\epsilon^{\dot{A}\dot{B}}$ are 
antisymmetric in the indices with 
$\epsilon^{12} = 1 = \epsilon^{\dot{1}\dot{2}}$,  
and $(\sigma^I)^A{}_B$ are components of the Pauli matrices. 
We use $\epsilon$'s to raise and lower indices, e.g.  
$(\sigma^I)^{AB} = \epsilon^{BC} (\sigma^I)^A{}_C$. 
For unitary representations the generators satisfy 
hermiticity conditions 
\begin{equation}
( L_n )^\dagger = L_{-n}, \qquad
( J_n^I )^\dagger = J_{-n}^I, \qquad
( G_r^{A\dot{A}} )^\dagger 
= \epsilon_{AB} \epsilon_{\dot{A}\dot{B}} G_{-r}^{B\dot{B}}. 
\label{hcond}
\end{equation}
The ${\cal N}=4$ super Virasoro algebra for the anti-holomorphic 
part has generators $\tilde{L}_n$, $\tilde{J}_n^{I'}$ and 
$\tilde{G}_r^{A'\dot{A}'}$ ($A'=1', 2'$; $\dot{A}'=\dot{1}', \dot{2}'$), 
which satisfy the similar (anti-)commutation relations and 
hermiticity conditions. 
\par
A chiral primary state $\ket{\phi_0}$ of the ${\cal N}=4$ 
superconformal algebra by definition satisfies 
\begin{eqnarray}
L_0 \ket{\phi_0} 
\A = \A j \ket{\phi_0}, \nonu 
J_0^3 \ket{\phi_0} 
\A = \A j \ket{\phi_0}, \nonu 
J_0^+ \ket{\phi_0} 
\A = \A 0, \nonu 
L_n \ket{\phi_0} \A = \A 0 \qquad (n > 0), \nonu
J_n^I \ket{\phi_0} \A = \A 0 \qquad (n > 0), \nonu
G_r^{A\dot{A}} \ket{\phi_0} \A = \A 0 \qquad (r > 0), \nonu
G_{-{1 \over 2}}^{2\dot{A}} \ket{\phi_0} \A = \A 0, 
\label{n4action}
\end{eqnarray}
where we have defined $J_0^\pm = J_0^1 \pm i J_0^2$. 
$L_0$ and $J_0^3$ must have the same eigenvalue 
$j = 0, {1 \over 2}, 1, {3 \over 2}, \cdots$. 
One can construct descendent states by applying other 
generators on $\ket{\phi_0}$. A chiral primary state and its 
descendent states are grouped into highest weight 
representations of the Virasoro and SU(2) Kac-Moody algebras.  
For $j \geq 1$ the corresponding highest weight states are 
\begin{eqnarray}
\ket{\phi_0} \A {} \A \nonu
\ket{\phi_1^{\dot{A}}} 
\A = \A G_{-{1 \over 2}}^{1\dot{A}} \ket{\phi_0}, \nonu
\ket{\phi_2} 
\A = \A \left( G_{-{1 \over 2}}^{12} G_{-{1 \over 2}}^{11} 
+ {1 \over 2j} L_{-1} J_0^- \right) \ket{\phi_0}. 
\label{short}
\end{eqnarray}
The second term in $\ket{\phi_2}$ is needed so that 
$\ket{\phi_2}$ becomes a highest weight state of the Virasoro 
and SU(2) Kac-Moody algebras. 
The eigenvalues of $L_0$ and $J_0^3$ for these states are 
\begin{eqnarray}
\begin{array}{ccc}
 \quad & \quad L_0 \quad & \quad J_0^3 \quad \\
\ket{\phi_0} \quad & j & j \\
\ket{\phi_1^{\dot{A}}} \quad & j+{1 \over 2} & j-{1 \over 2} \\
\ket{\phi_2} \quad & j+1 & j-1. 
\end{array}
\end{eqnarray}
For $j={1 \over 2}$ there exist only $\ket{\phi_0}$ and 
$\ket{\phi_1^{\dot{A}}}$, and no $\ket{\phi_2}$. 
\par
In refs.\ \cite{MS,DKSS,BOER,APT} the Kaluza-Klein spectrum of 
six-dimensional supergravities for the compactification 
on AdS${}_3$ $\times$ S${}^3$ 
was obtained and compared to the spectrum of chiral primary 
states of two-dimensional superconformal field theories. 
We can identify the perturbations (\ref{sd1})--(\ref{asd2}) 
in this spectrum. The Kaluza-Klein spectrum of the six-dimensional 
${\cal N}=(4,4)$ supergravity on AdS${}_3$ $\times$ S${}^3$ 
obtained in ref.\ \cite{BOER} is 
\begin{eqnarray}
\A\A \bigoplus_{{\bf m}={\bf 2}}^\infty \left[ 
({\bf m}, {\bf m+2})_S + ({\bf m+2}, {\bf m})_S 
+ 4 ({\bf m}, {\bf m+1})_S 
+ 4 ({\bf m+1}, {\bf m})_S \right] \nonu
\A\A {} \qquad + \bigoplus_{{\bf m}={\bf 3}}^\infty 
\left[ 6 ({\bf m}, {\bf m})_S \right]
+ 5 ({\bf 2}, {\bf 2})_S. 
\label{t4}
\end{eqnarray}
Here, $({\bf m}, {\bf m}')_S$ represents a short representation 
of the superalgebra SU($2|1,1$) $\times$ SU($2|1,1$). 
It is a product representation of two short representations 
${\bf m}_S$ and ${\bf m}'_S$ for each SU($2|1,1$). 
The superalgebra SU($2|1,1$) is a subalgebra of the 
${\cal N}=4$ superconformal algebra (\ref{n4algebra}) consisting of 
the SO(2,1) generators $L_0$, $L_{\pm1}$, the SU(2) generators $J_0^I$ 
and the supersymmetry generators $G_{\pm{1 \over 2}}^{A\dot{A}}$. 
The representation ${\bf m}_S$ consists of four irreducible 
representations of SO(2,1) $\times$ SU(2) whose highest weight 
states are given in eq.\ (\ref{short}) with $j={1 \over 2}(m-1)$. 
Namely, there is a one-to-one correspondence between 
a representation ${\bf m}_S$ and a chiral primary state 
with $j={1 \over 2}(m-1)$. 
\par
The conformal weights $(h, \bar{h})$ of the perturbations 
$G_2^{(+)}$ are determined by their spins and $r$-dependences. 
Since the perturbations (\ref{sd1})--(\ref{asd2}) are spin 0 scalars 
on the two-dimensional boundary of AdS${}_3$ we have $h = \bar{h}$. 
When $G_{ij}^{(+)} \sim r^s$ in the coordinate frame, we have 
$G_{\hat{i}\hat{j}}^{(+)} 
\sim r^s \times \left( Z^{-{1 \over 2}} \right)^2 
\sim r^{s+2}$ in the inertial (local Lorentz) frame. 
We obtain a relation $h + \bar h = d + (s+2) = s+4$ ($d=2$). 
Under the isometry SO(4) = SU(2) $\times$ SU(2) of S${^3}$ 
the self-dual and the anti self-dual two-forms in 
eqs.\ (\ref{sdt}) and (\ref{asdt}) transform as 
$({\bf 3}, {\bf 1})$ and $({\bf 1}, {\bf 3})$. 
Therefore, the perturbations $G_2^{(+)}$ in 
eqs.\ (\ref{sd1})--(\ref{asd2}) have the quantum numbers 
in Table \ref{table1}. 
\begin{table}[t]
\begin{center}
\renewcommand{\arraystretch}{1.3}
\begin{tabular}{|c||c|c|c|c|} \hline
perturbation & $(h, \bar{h})$ & 
SU(2) $\times$ SU(2) & multiplicity & supermultiplet \\ \hline
eq.\ (\ref{sd1}) & $(2,2)$ & $({\bf 3}, {\bf 1})$ & 8 
& $({\bf 3}, {\bf 4})_S$ \\
eq.\ (\ref{sd2}) & $(1,1)$ & $({\bf 3}, {\bf 1})$ & 8 
& $({\bf 3}, {\bf 2})_S$ \\
eq.\ (\ref{asd1}) & $(2,2)$ & $({\bf 1}, {\bf 3})$ & 8 
& $({\bf 4}, {\bf 3})_S$ \\
eq.\ (\ref{asd2}) & $(1,1)$ & $({\bf 1}, {\bf 3})$ & 8 
& $({\bf 2}, {\bf 3})_S$ \\ 
\hline
\end{tabular}
\end{center}
\caption{Perturbations.}
\label{table1}
\end{table}
The multiplicity is 8 since $G^{(+)\alpha\dot{\alpha}}_2$
has two internal indices $\alpha=1, ..., 4$, 
$\dot\alpha = 1, ..., 4$ and the projections 
${1 \over 2}(1 \pm \gamma_5)$ reduce the 16 components by half. 
Looking for short supermultiplets in eq.\ (\ref{t4}) which 
contain states having these quantum numbers we find that only 
supermultiplets shown in the last column of Table \ref{table1} 
contain those states. Explicitly, the perturbations correspond 
to the following states in the CFT  
\begin{eqnarray}
{\rm eq.}\ (\ref{sd1}): \A\A \quad 
\ket{\phi_1^{\dot{A}}(j=\textstyle{3 \over 2})} 
\otimes \ket{\tilde{\phi}_2(\bar{j}=1)}, \nonu
{\rm eq.}\ (\ref{sd2}): \A\A \quad 
\ket{\phi_1^{\dot{A}}(j=\textstyle{1 \over 2})} 
\otimes \ket{\tilde{\phi}_0(\bar{j}=1)}, \nonu
{\rm eq.}\ (\ref{asd1}): \A\A \quad \ket{\phi_2(j=1)} \otimes
\ket{\tilde{\phi}_1^{\dot{A}'}(\bar{j}=\textstyle{3 \over 2})}, \nonu
{\rm eq.}\ (\ref{asd2}): \A\A \quad \ket{\phi_0(j=1)} \otimes
\ket{\tilde{\phi}_1^{\dot{A}'}(\bar{j}=\textstyle{1 \over 2})}. 
\label{cftpert}
\end{eqnarray}
\par
To examine unbroken symmetries in the CFT side 
we need to know the action of supercharges on these states. 
{}From eqs.\ (\ref{short}), (\ref{n4action}), (\ref{n4algebra}) 
we obtain 
\begin{eqnarray}
G_{{1 \over 2}}^{2\dot{A}} \ket{\phi_0} 
\A = \A G_{{1 \over 2}}^{1\dot{A}} \ket{\phi_0} 
= G_{-{1 \over 2}}^{2\dot{A}} \ket{\phi_0} = 0, \nonu
G_{-{1 \over 2}}^{1\dot{A}} \ket{\phi_0} 
\A = \A \ket{\phi_1^{\dot{A}}}, \nonu
%
G_{{1 \over 2}}^{2\dot{A}} \ket{\phi_1^{\dot{B}}}
\A = \A -2j \epsilon^{\dot{A}\dot{B}} \ket{\phi_0}, \nonu
G_{{1 \over 2}}^{1\dot{A}} \ket{\phi_1^{\dot{B}}}
\A = \A \epsilon^{\dot{A}\dot{B}} J_0^- \ket{\phi_0}, \nonu
G_{-{1 \over 2}}^{2\dot{A}} \ket{\phi_1^{\dot{B}}}
\A = \A -\epsilon^{\dot{A}\dot{B}} L_{-1} \ket{\phi_0}, \nonu
G_{-{1 \over 2}}^{1\dot{A}} \ket{\phi_1^{\dot{b}}}
\A = \A - \epsilon^{\dot{A}\dot{B}} \left( \ket{\phi_2}
- {1 \over 2j} L_{-1} J_0^- \ket{\phi_0} \right), \nonu
%
G_{{1 \over 2}}^{2\dot{A}} \ket{\phi_2} 
\A = \A \left({1 \over 2j} - 2j \right) 
\ket{\phi_1^{\dot{A}}}, \nonu
G_{{1 \over 2}}^{1\dot{A}} \ket{\phi_2}
\A = \A \left( 1 +{1 \over 2j} \right) J_0^- 
\ket{\phi_1^{\dot{A}}}, \nonu
G_{-{1 \over 2}}^{2\dot{A}} \ket{\phi_2}
\A = \A - \left( 1 - {1 \over 2j} \right) L_{-1} 
\ket{\phi_1^{\dot{A}}}, \nonu
G_{-{1 \over 2}}^{1\dot{A}} \ket{\phi_2}
\A = \A {1 \over 2j} L_{-1} J_0^- \ket{\phi_1^{\dot{A}}}. 
\label{gonphi}
\end{eqnarray}
We can express these relations in terms of local operators 
and currents. For each state we introduce a local operator 
$\phi(z)$ which create the state as 
\begin{equation}
\ket{\phi} = \phi(0) \ket{0}, 
\end{equation}
where $\ket{0}$ is the SU($2|1,1$) invariant vacuum. We also 
introduce currents for the ${\cal N}=4$ superconformal generators 
\begin{eqnarray}
T(z) \A = \A \sum_{n \in {\bf Z}} z^{-n-2} L_n, \nonu
G^{A\dot{A}}(z) \A = \A \sum_{r \in {\bf Z}+{1 \over 2}} 
z^{-r-{3 \over 2}} G_r^{A\dot{A}}, \nonu 
J^I(z) \A = \A \sum_{n \in {\bf Z}} z^{-n-1} J_n^I. 
\end{eqnarray}
Then, the relations (\ref{gonphi}) lead to the OPEs
\begin{eqnarray}
G^{2\dot{A}}(z_1) \phi_0(z_2) \A \sim \A {\rm regular}, \nonu
G^{1\dot{A}}(z_1) \phi_0(z_2) 
\A \sim \A {1 \over z_1-z_2} \phi_1^{\dot{A}}(z_2), \nonu
G^{2\dot{A}}(z_1) \phi_1^{\dot{B}}(z_2) 
\A \sim \A - {2j \over (z_1-z_2)^2} 
\epsilon^{\dot{A}\dot{B}} \phi_0(z_2) 
- {1 \over z_1-z_2} \epsilon^{\dot{A}\dot{B}} 
\partial\phi_0(z_2), \nonu
G^{1\dot{A}}(z_1) \phi_1^{\dot{B}}(z_2) 
\A \sim \A {1 \over (z_1-z_2)^2} \epsilon^{\dot{A}\dot{B}} 
[ J_0^-, \phi_0(z_2) ] 
- {1 \over z_1-z_2} \epsilon^{\dot{A}\dot{B}} \left( \phi_2 
- {1 \over 2j} \partial [ J_0^-, \phi_0] \right)(z_2), \nonu
G^{2\dot{A}}(z_1) \phi_2(z_2) 
\A \sim \A  - \left( 2j - {1 \over 2j} \right) 
{1 \over (z_1-z_2)^2} \phi_1^{\dot{A}}(z_2) 
- \left( 1 - {1 \over 2j} \right) {1 \over z_1-z_2} 
\partial\phi_1^{\dot{A}}(z_2), \nonu
G^{1\dot{A}}(z_1) \phi_2(z_2) 
\A \sim \A  \left( 1 + {1 \over 2j} \right) 
{1 \over (z_1-z_2)^2} [ J_0^-, \phi_1^{\dot{A}}(z_2) ] 
+ {1 \over 2j} {1 \over z_1-z_2} \partial [ J_0^-, 
\phi_1^{\dot{A}}(z_2) ]. 
\label{ope}
\end{eqnarray}
{}From these OPEs one can obtain commutators 
of the generators $L_n$, $J_n^I$, $G_r^{A\dot{A}}$ and the local 
operators $\phi_0(z)$, $\phi_1^{\dot{A}}(z)$, $\phi_2(z)$ by 
computing contour integrals of $z_1$ around $z_2$. 
\par
The operators corresponding to the perturbations in the 
supergravity solution are integrated operators. 
There is an arbitrariness in the choice of the integration 
measure. Since the perturbation in the supergravity side is 
invariant under translations of $x^\mu$, we choose 
the measure invariant under the translation of $w$ 
in eq.\ (\ref{zw2}) (Euclidean version of $x^\mu$). 
Thus, we consider the integrated operators 
\begin{eqnarray}
\Phi_0 \A = \A \int d^2w \, \phi_0(w) \tilde{\phi}(\bar{w}) 
= \int d^2z \, \left| \textstyle{1 \over 2}(z+1)^2 \right|^{2(j-1)} 
\, \phi_0(z) \tilde{\phi}(\bar{z}), \nonu
\Phi_1^{\dot{A}} \A = \A \int d^2w \, \phi_1^{\dot{A}}(w) 
\tilde{\phi}(\bar{w}) 
= \int d^2z \, \left| \textstyle{1 \over 2}(z+1)^2 
\right|^{2(j-{1 \over 2})} \, \phi_1^{\dot{A}}(z) 
\tilde{\phi}(\bar{z}), \nonu
\Phi_2 \A = \A \int d^2w \, \phi_2(w) \tilde{\phi}(\bar{w}) 
= \int d^2z \, \left| \textstyle{1 \over 2}(z+1)^2 \right|^{2j} 
\, \phi_2(z) \tilde{\phi}(\bar{z}), 
\label{integrated}
\end{eqnarray}
where we have used the transformation property of the Virasoro 
primary field of conformal weight $h$: $\phi(w) 
= \left({\partial z \over \partial w}\right)^h \phi(z)$. 
In eq.\ (\ref{integrated}) we have specified only the holomorphic 
part of the local operators. The anti-holomorphic part 
$\tilde{\phi}$ can be $\tilde{\phi}_0$, $\tilde{\phi}_1^{\dot{A}'}$, 
$\tilde{\phi}_2$ with the same conformal weight as 
the holomorphic part. 
Using the fact that the local operators $\phi_0$, $\phi_1^{\dot{A}}$, 
$\phi_2$ and $\tilde{\phi}$ are primary fields of the Virasoro 
algebra it is easy to see that these integrated operators indeed 
commute with the translation generators of $w$, $\bar{w}$
\begin{equation}
P = L_{-1} + 2 L_0 + L_1, \qquad 
\tilde{P} = \tilde{L}_{-1} + 2 \tilde{L}_0 + \tilde{L}_1. 
\label{translation}
\end{equation}
By integrating the commutation relations between the generators 
and the local operators we obtain 
\begin{eqnarray}
[ G_r^{2\dot{A}}, \Phi_0 ] \A = \A 0, \nonu
[ G_r^{1\dot{A}}, \Phi_0 ] 
\A = \A \int d^2 z \left| \textstyle{1 \over 2}(z+1)^2 \right|^{2(j-1)} 
z^{r+{1 \over 2}} \phi_1^{\dot{A}}(z) \tilde{\phi}(\bar{z}), \nonu
[ G_r^{2\dot{A}}, \Phi_1^{\dot{B}} ] 
\A = \A -2 \left( j-\textstyle{1 \over 2} \right) 
\epsilon^{\dot{A}\dot{B}} \int d^2 z \left| \textstyle{1 \over 2}(z+1)^2 
\right|^{2(j-{1 \over 2})} {z^{r-{1 \over 2}} \over z+1} 
\left[ (r-\textstyle{1 \over 2})z + (r+\textstyle{1 \over 2}) 
\right] \nonu
\A\A \times \phi_0(z) \tilde{\phi}(\bar{z}), \nonu
[ G_r^{1\dot{A}}, \Phi_1^{\dot{B}} ] 
\A = \A {1 \over j} \left( j-\textstyle{1 \over 2} \right) 
\epsilon^{\dot{A}\dot{B}} \int d^2 z \left| \textstyle{1 \over 2}(z+1)^2 
\right|^{2(j-{1 \over 2})} {z^{r-{1 \over 2}} \over z+1} 
\left[ (r-\textstyle{1 \over 2})z + (r+\textstyle{1 \over 2}) 
\right] \nonu 
\A\A \times [ J_0^-, \phi_0(z) ] \tilde{\phi}(\bar{z})
- \epsilon^{\dot{A}\dot{B}} \int d^2z \left| 
\textstyle{1 \over 2}(z+1)^2 \right|^{2(j-{1 \over 2})} 
z^{r+\textstyle{1 \over 2}} \phi_2(z) \tilde{\phi}(\bar{z}), \nonu
[ G_r^{2\dot{A}}, \Phi_2 ] 
\A = \A - 2 \left( j-\textstyle{1 \over 2} \right) 
\int d^2 z \left| \textstyle{1 \over 2}(z+1)^2 
\right|^{2j} {z^{r-{1 \over 2}} \over z+1} 
\left[ (r-\textstyle{1 \over 2})z + (r+\textstyle{1 \over 2}) 
\right] \phi_1^{\dot{A}}(z) \tilde{\phi}(\bar{z}), \nonu
[ G_r^{1\dot{A}}, \Phi_2 ] 
\A = \A \int d^2 z \left| \textstyle{1 \over 2}(z+1)^2 
\right|^{2j} {z^{r-{1 \over 2}} \over z+1} 
\left[ (r-\textstyle{1 \over 2})z + (r+\textstyle{1 \over 2}) 
\right] [ J_0^-, \phi_1^{\dot{A}}(z) ] \tilde{\phi}(\bar{z}). 
\label{intcommutator}
\end{eqnarray}
\par
Let us find unbroken supersymmetries by the perturbations. 
The $(h,\bar{h})=(2,2)$ operators corresponding to the first and 
third states in eq.\ (\ref{cftpert}) are 
\begin{eqnarray}
\Phi_{\dot{A}} \A = \A \int d^2w \, 
\phi_{1 \dot{A}}(w; j=\textstyle{3 \over 2}) 
\, \tilde{\phi}_2(\bar{w}; \bar{j}=1), \nonu
\Phi_{\dot{A}'} \A = \A \int d^2w \, \phi_2(w; j=1) \, 
\tilde{\phi}_{1 \dot{A}'}(\bar{w}; \bar{j}=\textstyle{3 \over 2}). 
\label{22}
\end{eqnarray}
Here, we have lowered the indices $\dot{A}$, $\dot{A}'$ by using 
$\epsilon_{\dot{A}\dot{B}}$, $\epsilon_{\dot{A}'\dot{B}'}$. 
{}From eq.\ (\ref{intcommutator}) we find that the supercharges 
which commute with $\Phi_{\dot{A}'}$ for a given $\dot{A}'$ are 
\begin{equation}
G_{-{1 \over 2}}^{B\dot{B}} 
+ G_{1 \over 2}^{B\dot{B}}, \qquad
\tilde{G}_{-{1 \over 2}}^{2'\dot{A}'} 
+ \tilde{G}_{1 \over 2}^{2'\dot{A}'}, \qquad
\tilde{G}_{\pm{1 \over 2}}^{B'\dot{B}'}, 
\label{unbrokensusy1}
\end{equation}
where $B$, $B'$, $\dot{B}$ are arbitrary and 
$\dot{B}'\not=\dot{A}'$. 
The bosonic generators which commute with $\Phi_{\dot{A}'}$ are 
$P$, $\tilde{P}$ and $J_0^I$. To add these operators 
to the CFT Hamiltonian as a perturbation 
we should make a hermitian combination 
$m \Phi_{\dot{A}'} + m^* \Phi_{\dot{A}'}^\dagger$, 
where $m$ is a complex constant. 
The supersymmetries preserved by this perturbation are those 
preserved by both of $\Phi_{\dot{A}'}$ and $\Phi_{\dot{A}'}^\dagger$. 
Remembering the hermiticity condition (\ref{hcond}) we find that 
the unbroken supercharges are 
\begin{equation}
Q^{B\dot{B}} 
= G_{-{1 \over 2}}^{B\dot{B}} + G_{1 \over 2}^{B\dot{B}}, \qquad
\tilde{Q} 
= \tilde{G}_{-{1 \over 2}}^{2'\dot{B}'} 
+ \tilde{G}_{1 \over 2}^{2'\dot{B}'},  
\label{unbrokensusy2}
\end{equation}
where $B$, $\dot{B}$ are arbitrary and $\dot{B}'\not=\dot{A}'$. 
These supercharges together with the translation generators 
(\ref{translation}) satisfy the ${\cal N}=(4,2)$ Poincar\'e 
supersymmetry algebra 
\begin{equation}
\{ Q^{A\dot{A}}, Q^{B\dot{B}} \} 
= \epsilon^{AB} \epsilon^{\dot{A}\dot{B}} P, \qquad 
\{ \tilde{Q}, \tilde{Q}^\dagger \} = \tilde{P}. 
\end{equation}
The supersymmetries preserved by $\Phi_{\dot{A}}$ in 
eq.\ (\ref{22}) and its hermitian conjugate are similarly 
obtained and form the ${\cal N}=(2,4)$ Poincar\'e 
supersymmetry algebra. 
\par
The $(h,\bar{h})=(1,1)$ operators corresponding to the second and 
fourth states in eq.\ (\ref{cftpert}) are 
\begin{eqnarray}
\Phi_{\dot{A}} \A = \A \int d^2w \, 
\phi_{1 \dot{A}}(w; j=\textstyle{1 \over 2}) \, 
\tilde{\phi}_0(\bar{w}; \bar{j}=1), \nonu
\Phi_{\dot{A}'} \A = \A \int d^2w \, \phi_0(w; j=1) \, 
\tilde{\phi}_{1 \dot{A}'}(\bar{w}; \bar{j}=\textstyle{1 \over 2}). 
\label{11}
\end{eqnarray}
{}From eq.\ (\ref{intcommutator}) the supercharges which 
commute with $\Phi_{\dot{A}'}$ for a given $\dot{A}'$ are 
\begin{equation}
G_{\pm{1 \over 2}}^{1\dot{B}}, \qquad
\tilde{G}_{\pm{1 \over 2}}^{B'\dot{B}'}, 
\label{unbrokensusy3}
\end{equation}
where $B'$, $\dot{B}$, $\dot{B}'$ are arbitrary. 
The bosonic generators which commute with $\Phi_{\dot{A}'}$ 
are $P$, $\tilde{L}_{\pm 1}$, $\tilde{L}_0$ 
and $\tilde{J}_0^{I'}$. 
The supercharges which commute with a hermitian perturbation 
$m \Phi_{\dot{A}'} + m^* \Phi_{\dot{A}'}^\dagger$ are 
\begin{eqnarray}
\tilde{G}_{\pm{1 \over 2}}^{B'\dot{B}'},
\label{unbrokensusy4} 
\end{eqnarray}
which together with $\tilde{L}_{-1}$, $\tilde{L}_0$, 
$\tilde{L}_1$ and $\tilde{J}_0^{I'}$ form 
the ${\cal N}=(0,4)$ superconformal algebra. 
The supersymmetries preserved by $\Phi_{\dot{A}}$ and its 
hermitian conjugate are similarly obtained and form the 
${\cal N}=(4,0)$ superconformal algebra. 
\par
In the next section we will show that the perturbations in 
the supergravity solution preserve the same supersymmetries 
as above. 
\par
%
\newsection{Unbroken symmetries by perturbations in supergravity}
%
Let us start from the bosonic symmetries, i.e., the isometry 
of AdS${}_3$ $\times$ S${}^3$. 
The coordinate transformation of the perturbation $G_2$ 
for $\delta x^M = \xi^M$ is 
\begin{equation}
\delta G_{MN} = \xi^P \partial_P G_{MN} 
+ \partial_M \xi^P G_{PN} + \partial_N \xi^P G_{MP}. 
\label{coordinatetransformation}
\end{equation}
Our perturbations (\ref{sd1})--(\ref{asd2}) have only non-vanishing 
components $G_{ij}$ and are independent of $x^\mu$. 
\par
In the case of the Killing vectors (\ref{killing}) of AdS${}_3$ 
this transformation gives 
\begin{eqnarray}
\delta G_{\mu\nu} 
\A = \A 0, \nonu
\delta G_{\mu i} 
\A = \A - {1 \over 2} \partial_\mu \partial_\rho \zeta^\rho x^j 
G_{ji}, \nonu
\delta G_{ij}
\A = \A - {1 \over 2} \partial_\rho \zeta^\rho 
\left( r \partial_r + 2 \right) G_{ij}. 
\end{eqnarray}
For Poincar\'e transformations, for which 
$\partial_\rho \zeta^\rho = 0$, they automatically vanish. 
However, the invariance under all of the Killing vectors 
(\ref{killing}) requires 
\begin{equation}
x^i G_{ij} = 0, \qquad
\left( r \partial_r + 2 \right) G_{ij} = 0, 
\label{conformalg}
\end{equation}
namely, $G_{ij}$ should lie along the S${}^3$ directions and have 
$r$-dependence $r^{-2}$. These conditions are satisfied by the 
perturbations $G^{(+)}_2$ in eqs.\ (\ref{sd2}), (\ref{asd2}) but 
not by those in eqs.\ (\ref{sd1}), (\ref{asd1}). 
This is consistent with the result in the CFT side that 
local operators corresponding to (\ref{sd2}), (\ref{asd2}) have 
conformal weight $(1,1)$ and the integrated operators are 
invariant under the conformal transformations 
while those corresponding to (\ref{sd1}), 
(\ref{asd1}) are only invariant under the translations. 
\par
The Killing vectors for the isometry SO(3) $\sim$ 
SU(2) $\times$ SU(2) of S${}^3$ are $\xi^i = \lambda^i{}_j x^j$, 
where $\lambda_{ij}=-\lambda_{ji}$. The net effect of the 
transformation (\ref{coordinatetransformation}) in this case 
is that the components $T_{ij}$ in $G_2$ are changed according 
to the SO(3) transformation. Since $T_{ij}$ belong to 
a representation $({\bf 3},{\bf 1})$ or $({\bf 1},{\bf 3})$, 
SU(2) $\times$ SU(2) symmetry is broken to 1 $\times$ SU(2) or 
SU(2) $\times$ 1. 
\par
Finally, let us obtain supersymmetries preserved by the 
perturbations. 
We consider the case in which only the $m_1$ terms in 
eqs.\ (\ref{sdt}), (\ref{asdt}) are present. These terms 
represent the highest weight state of $({\bf 3},{\bf 1})$ 
or $({\bf 1},{\bf 3})$ of SU(2) $\times$ SU(2), and corresponds 
to the CFT operator in eqs.\ (\ref{22}), (\ref{11}). 
Therefore, we should recover unbroken supersymmetries in 
eqs.\ (\ref{unbrokensusy1}), (\ref{unbrokensusy3}). 
\par
We shall obtain transformation parameter 
$\epsilon$ for which the supertransformations of the fermionic 
fields (\ref{susytransformation}) vanish to the first order 
in $G_2$. 
First, the condition $\delta \chi_{+a\dot{\alpha}} = 0$ require 
\begin{equation}
\delta \chi_{+a\dot{\alpha}} 
= {1 \over R} S_a \hat{\gamma}^{\hat{r}} \bar\gamma_{2D} 
\epsilon_{-\dot{\alpha}}
- {1 \over 4} G_{ij\alpha\dot{\alpha}} \hat{\gamma}^{ij} 
\epsilon_+^\beta (\gamma_a)_\beta{}^\alpha = 0. 
\end{equation}
Multiplying $S^a$ to this equation we can express $\epsilon_-$ 
in terms of $\epsilon_+$ as 
\begin{equation}
\epsilon_{-\dot{\alpha}}
= {1 \over 4} R \, G_{ij\alpha\dot{\alpha}} \, 
\hat{\gamma}^{\hat{r}} \hat{\gamma}^{ij} \bar\gamma_{2D} 
\epsilon_+^{\beta} (\gamma_5)_\beta{}^\alpha. 
\label{epsilonminus}
\end{equation}
We see that $\epsilon_-$ is non-vanishing and of order $G_2$. 
The condition $\delta\psi_{+M\alpha}=0$ gives the 
same condition as the unperturbed background (\ref{killingspinor}) 
to the first order in $G_2$, 
whose solution is eq.\ (\ref{susyparameter}). 
The condition $\delta\chi_{-\dot{a}\alpha}=0$ is also 
automatically satisfied. 
Substituting eq.\ (\ref{epsilonminus}) into 
$\delta \psi_{-M\dot{\alpha}}$ and $\delta \chi_{+a\dot{\alpha}}$ 
and using the differential equation on $\epsilon_+$ 
(\ref{killingspinor2}) we obtain 
\begin{eqnarray}
\delta \psi_{-\mu}{}^{\dot{\alpha}} 
\A = \A {1 \over 8} G_{kl}^{\alpha\dot{\alpha}} 
\hat{\gamma}_\mu \bar\gamma_{4D} 
[ \hat{\gamma}^{\hat{r}}, \hat{\gamma}^{kl} ]
\left( \delta_\alpha^\beta - \bar{\gamma}_{4D} 
(\gamma_5)_\alpha{}^\beta \right) \epsilon_{+\beta}, \nonu
\delta \psi_{-i}{}^{\dot{\alpha}} 
\A = \A {1 \over 8} G_{kl}{}^{\alpha\dot{\alpha}}
\left( \hat{\gamma}_i \hat{\gamma}^{kl} 
- \hat{\gamma}^{\hat{r}} \hat{\gamma}^{kl} 
\hat{\gamma}^{\hat{r}} \hat{\gamma}_i \right) 
\left( \delta_\alpha^\beta - \bar{\gamma}_{4D} 
(\gamma_5)_\alpha{}^\beta \right) \epsilon_{+\beta} \nonu
\A\A - {R \over 4r^2} \partial_i ( r^2 
G_{kl}{}^{\alpha\dot{\alpha}} ) 
\hat{\gamma}^{\hat{r}} \hat{\gamma}^{kl} 
\bar{\gamma}_{4D} (\gamma_5)_\alpha{}^\beta \epsilon_{+\beta} 
- G_{ik}{}^{\alpha\dot{\alpha}} \hat{\gamma}^k 
\epsilon_{+\alpha}, \nonu 
\delta \chi_{+a}{}^{\dot{\alpha}} 
\A = \A {1 \over 4} G_{ij}{}^{\alpha\dot{\alpha}} 
\hat{\gamma}^{ij} (\gamma_a)_\alpha{}^\beta 
\epsilon_{+\beta} \qquad (a=1,2,3,4). 
\label{susycondition}
\end{eqnarray}
Supersymmetry parameters for which these transformations 
vanish correspond to unbroken supersymmetries. 
We examine the conditions (\ref{susycondition}) for each 
of the perturbations (\ref{sd1})--(\ref{asd2}). 
\par
First, let us consider the perturbations $G^{(+)}_2$ in 
eq.\ (\ref{sd1}). In this case $G^{(+)}_{ij}$ is constant and 
self-dual. The self-duality implies 
\begin{equation}
G^{(+)}_{kl} \hat{\gamma}^{kl} \psi = 0 
\label{sdid}
\end{equation}
when $\bar{\gamma}_{4D} \psi = \psi$. 
We first consider $\delta\chi_+$. 
When $\epsilon_+$ satisfies $\gamma_5 \epsilon_+ = -\epsilon_+$ 
or $\bar{\gamma}_{4D} \epsilon_+ = \epsilon_+$, it vanishes 
because of $G^{(+)}_2 \gamma_5 = - G^{(+)}_2$ or 
the identity (\ref{sdid}). 
When $\bar{\gamma}_{4D} \epsilon_+ = - \epsilon_+$ and 
$\gamma_5 \epsilon_+ = \epsilon_+$, it is proportional to 
\begin{equation}
(m_1)^{\alpha\dot{\alpha}} \hat{\gamma}^{1\bar{2}} 
(\gamma_a)_\alpha{}^\beta \epsilon_{+\beta}. 
\end{equation}
By substituting $\epsilon_+^{(-+)}$ in eq.\ (\ref{susyparameter}) 
into this equation 
and using the explicit form of gamma matrices $\hat{\gamma}^i$ 
and $\gamma^a$ in Appendix we find that it does not vanish. 
Therefore, $\delta\chi_+=0$ requires $\epsilon_{1A\dot{A}}^{(++)}=0$. 
As for $\delta\psi_{+\mu}$, it vanishes when 
$\bar{\gamma}_{4D} \epsilon_+ = - \epsilon_+$ or 
$\gamma_5 \epsilon_+ = \epsilon_+$ as seen from 
eq.\ (\ref{susycondition}). 
When $\bar{\gamma}_{4D} \epsilon_+ = \epsilon_+$ and 
$\gamma_5 \epsilon_+ = - \epsilon_+$, $\delta\psi_{+\mu}$ is 
proportional to 
\begin{equation}
(m_1)^{\alpha\dot{\alpha}} \left( 
z^1 \hat{\gamma}^{\bar{2}} - \bar{z}^2 \hat{\gamma}^1 
\right) \epsilon_{+\alpha}. 
\end{equation}
Substituting $\epsilon_+^{(+-)}$ in eq.\ (\ref{susyparameter}) 
we find that $\delta\psi_{+\mu}=0$ requires 
$\epsilon_{11'\dot{2}'}^{(--)}=0$ when 
$(m_1)^{\dot{1}'\dot{\alpha}} =0$, 
and $\epsilon_{11'\dot{1}'}^{(--)}=0$ 
when $(m_1)^{\dot{2}'\dot{\alpha}} =0$. 
Here, we have used two-component spinor index 
$\dot{A}'=\dot{1}', \dot{2}'$ instead of the four-component one 
$\alpha = 1,2,3,4$ for the first index of 
$(m_1)^{\alpha\dot{\alpha}}$ (See Appendix.). 
Finally, let us consider $\delta\psi_{-i}$. 
When $\gamma_5 \epsilon_+ = \epsilon_+$, it automatically vanishes. 
When $\gamma_5 \epsilon_+ = - \epsilon_+$, the condition 
$\delta\psi_{-i}=0$ is shown to be equivalent to 
$G_{ij}^{(+)\alpha\dot{\alpha}} \hat{\gamma}^j \epsilon_{+\alpha} 
= 0$, i.e., 
\begin{equation}
(m_1)^{\alpha\dot{\alpha}} \hat{\gamma}^1 \epsilon_{+\alpha} 
= 0, \qquad 
(m_1)^{\alpha\dot{\alpha}} \hat{\gamma}^{\bar{2}} 
\epsilon_{+\alpha} = 0. 
\end{equation}
Substituting $\epsilon_+^{(\pm-)}$ in eq.\ (\ref{susyparameter}) 
we find that $\delta\psi_{+i}=0$ requires 
$\epsilon_{01'\dot{2}'}^{(--)}=0$, $\epsilon_{1A'\dot{2}'}^{(--)}=0$
when $(m_1)^{\dot{1}'\dot{\alpha}} =0$, 
and $\epsilon_{01'\dot{1}'}^{(--)}=0$, $\epsilon_{1A'\dot{1}'}^{(--)}=0$
when $(m_1)^{\dot{2}'\dot{\alpha}}=0$. 
To summarize, the unbroken supersymmetries by the perturbation, 
for which all of the transformations in eq.\ (\ref{susycondition}) 
vanish are given as follows. When only 
$G_2^{(+)\dot{A}'\dot{\alpha}}$ 
is non-vanishing for a given $\dot{A}'$, the transformation 
parameters of the unbroken supersymmetries are 
\begin{equation}
\epsilon_{0B\dot{B}}^{(++)}, \quad
\epsilon_{02'\dot{A}'}^{(--)}, \quad
\epsilon_{0B'\dot{B}'}^{(--)}, \quad
\epsilon_{1B'\dot{B}'}^{(--)}, 
\end{equation}
where $B$, $B'$, $\dot{B}$ are arbitrary 
and $\dot{B}' \not= \dot{A}'$. 
Remembering the correspondence between the supersymmetry 
parameters and the supercharges discussed at the end of 
sect.\ 2 we see that this result is in complete 
agreement with the unbroken 
supersymmetries in the CFT (\ref{unbrokensusy1}). 
\par
Next, let us consider the perturbations $G^{(+)}_2$ in 
eq.\ (\ref{asd2}). In this case $G^{(+)}_2$ satisfies 
eq.\ (\ref{conformalg}). We easily find that 
$\delta \psi_{-\mu} = 0$ for an arbitrary $\epsilon_+$ since 
\begin{equation}
G^{(+)}_{kl} [ \hat{\gamma}^{\hat{r}}, \hat{\gamma}^{kl} ]
= {4 \over R} x^k G^{(+)}_{kl} \hat{\gamma}^l 
\end{equation}
vanishes because of eq.\ (\ref{conformalg}). 
As for $\delta\psi_{-i}$ we need a formula 
for derivative of $G_{kl}^{(+)}$. From the explicit form of 
$G_{kl}^{(+)}$ in eq.\ (\ref{asd2}) we find 
\begin{equation}
\partial_i (r^2 G^{(+)}_{kl}) 
= - x^k G^{(+)}_{il} 
+ \left( \delta_{ik} - {x^i x^k \over r^2} 
\right) x^j (*_4 G^{(+)})_{jl} - (k \leftrightarrow l). 
\end{equation}
Substituting this into $\delta\psi_{-i}$ in 
eq.\ (\ref{susycondition}) and doing some algebra we find 
that $\delta\psi_{-i} = 0$ for an arbitrary $\epsilon_+$. 
Non-trivial conditions on $\epsilon_+$ only come from 
$\delta\chi_+ = 0$. When $\gamma_5 \epsilon_+ = - \epsilon_+$, 
$\delta\chi_+$ automatically vanishes. When 
$\gamma_5 \epsilon_+ = \epsilon_+$, we find that $\delta\chi_+$ 
vanishes only if $G^{(+)}_{ij} \hat{\gamma}^{ij} \epsilon_+ = 0$. 
Substituting $\epsilon_+^{(\pm+)}$ in eq.\ (\ref{susyparameter}) 
into this equation and using the explicit form 
\begin{equation}
G^{(+)}_{ij} \hat{\gamma}^{ij} 
= {1 \over 2} \alpha r^{-2} m_1 (1-\gamma_5) 
\left[ {1 \over 2} \hat{\gamma}^{12} 
- {z^1 z^2 \over r^2} (\hat{\gamma}^{1\bar{1}}  
- \hat{\gamma}^{2\bar{2}}) 
- {(z^2)^2 \over r^2} \hat{\gamma}^{1\bar{2}} 
- {(z^1)^2 \over r^2} \hat{\gamma}^{\bar{1}2} 
\right] 
\end{equation}
we find that $\delta\chi_+=0$ requires 
$\epsilon_{02\dot{A}}^{(++)}=0$, 
$\epsilon_{12\dot{A}}^{(++)}=0$. Thus, the unbroken 
supersymmetries for the perturbation (\ref{asd2}) are 
\begin{equation}
\epsilon_{01\dot{B}}^{(++)}, \quad
\epsilon_{11\dot{B}}^{(++)}, \quad
\epsilon_{0B'\dot{B}'}^{(--)}, \quad
\epsilon_{1B'\dot{B}'}^{(--)}, 
\end{equation}
where $B'$, $\dot{B}$, $\dot{B}'$ are arbitrary. 
This result is again in complete agreement with the unbroken 
supersymmetries in the CFT (\ref{unbrokensusy3}). 
\par
%

\vspace{10mm}
\noindent {\Large{\bf Acknowledgements}} 
\vspace{3mm}

One of the authors (M.N.) would like to thank 
the Center for Theoretical Physics of MIT, 
Department of Physics, University of Waterloo, 
and the Perimeter Institute for hospitality. 
The work of M.N. is supported in part by the Wenner-Gren Foundation. 
The work of Y.T. is supported in part by the Grant-in-Aid for 
Scientific Research from the Ministry of Education, 
Culture, Sports, Science and Technology, Japan, No.\ 15540252. 
%
%
\def\numberbysectiona{\@addtoreset{equation}{section}
\def\theequation{A.\arabic{equation}}}
\numberbysectiona
\vspace{7mm}
\pagebreak[3]
\setcounter{section}{1}
\setcounter{equation}{0}
\setcounter{subsection}{0}
\setcounter{footnote}{0}
\begin{center}
{\large{\bf Appendix: SO(4) and SO(5) gamma matrices}}
\end{center}
\nopagebreak
\medskip
\nopagebreak
\hspace{3mm}
Our representation of the SO(4) gamma matrices is 
\begin{eqnarray}
\hat{\gamma}^{\hat{2}} \A = \A \left( 
\begin{array}{cc}
0 & -i \\
i & 0
\end{array}
\right), \quad
\hat{\gamma}^{\hat{3}} = \left( 
\begin{array}{cc}
0 & -\sigma^2 \\
-\sigma^2 & 0
\end{array}
\right), \nonu
\hat{\gamma}^{\hat{4}} \A = \A \left( 
\begin{array}{cc}
0 & -\sigma^3 \\
-\sigma^3 & 0
\end{array}
\right), \quad
\hat{\gamma}^{\hat{5}} = \left( 
\begin{array}{cc}
0 & \sigma^1 \\
\sigma^1 & 0
\end{array}
\right), 
\end{eqnarray}
where $\sigma^1$, $\sigma^2$, $\sigma^3$ are 
the $2 \times 2$ Pauli matrices. 
The chirality matrix in eq.\ (\ref{chiralitymatrix}) then becomes 
\begin{equation}
\bar{\gamma}_{4D} 
= \hat{\gamma}^{\hat{2}} \hat{\gamma}^{\hat{3}} 
\hat{\gamma}^{\hat{4}} \hat{\gamma}^{\hat{5}} 
= \left( 
\begin{array}{cc}
1 & 0 \\
0 & -1
\end{array}
\right).
\end{equation}
An SO(4) spinor $\psi$ has components 
\begin{equation}
\psi
= \left( 
\begin{array}{c}
\psi_A \\
\psi_{A'} 
\end{array}
\right) \qquad
(A = 1,2;\ A' = 1',2'). 
\end{equation}
\par
Our representation of the SO(5) gamma matrices 
$(\gamma^a)_\alpha{}^\beta$ is 
\begin{eqnarray}
\gamma^1 \A = \A \left( 
\begin{array}{cc}
0 & \sigma^1 \\
\sigma^1 & 0
\end{array}
\right), \quad
\gamma^2 = \left( 
\begin{array}{cc}
0 & \sigma^2 \\
\sigma^2 & 0
\end{array}
\right), \quad
\gamma^3 = \left( 
\begin{array}{cc}
0 & \sigma^3 \\
\sigma^3 & 0
\end{array}
\right), \nonu
\gamma^4 \A = \A \left( 
\begin{array}{cc}
0 & -i \\
i & 0
\end{array}
\right), \quad
\gamma^5
= \left( 
\begin{array}{cc}
1 & 0 \\
0 & -1
\end{array}
\right). 
\end{eqnarray}
A four-component SO(5) spinor $\psi_\alpha$ ($\alpha=1,2,3,4$) is 
decomposed into two two-component spinors 
\begin{equation}
\psi
= \left( 
\begin{array}{c}
\psi_{\dot{A}} \\
\psi_{\dot{A}'} 
\end{array}
\right) \qquad
(\dot{A} = \dot{1}, \dot{2};\ \dot{A}' = \dot{1}', \dot{2}'). 
\end{equation}
The SO(5) charge conjugation matrix $\Omega^{\alpha\beta}$ satisfies 
\begin{equation}
\Omega \gamma^a \Omega^{-1} = (\gamma^a)^T
\end{equation}
and is given by 
\begin{equation}
\Omega 
= \left( 
\begin{array}{cc}
\epsilon^{\dot{A}\dot{B}} & 0 \\
0 & \epsilon^{\dot{A}'\dot{B}'}
\end{array}
\right), 
\end{equation}
where antisymmetric $\epsilon^{\dot{A}\dot{B}}$ and 
$\epsilon^{\dot{A}'\dot{B}'}$ are chosen as 
$\epsilon^{\dot{1}\dot{2}} = +1 = \epsilon^{\dot{1}'\dot{2}'}$. 
We also use antisymmetric $\epsilon_{\dot{A}\dot{B}}$ and 
$\epsilon_{\dot{A}'\dot{B}'}$ with 
$\epsilon_{\dot{1}\dot{2}} = +1 = \epsilon_{\dot{1}'\dot{2}'}$. 
\par
In eq.\ (\ref{susyparameter}) the supertransformation parameter 
$\epsilon_+$ is decomposed according to eigenvalues of 
$\bar{\gamma}_{4D}$ and $\gamma_5$. Each of the components 
has indices as 
\begin{equation}
\epsilon^{(++)}_{+A\dot{A}}, \qquad
\epsilon^{(+-)}_{+A\dot{A}'}, \qquad 
\epsilon^{(-+)}_{+A'\dot{A}}, \qquad 
\epsilon^{(--)}_{+A'\dot{A}'}. 
\end{equation}
%
%
\newcommand{\NP}[1]{{\it Nucl.\ Phys.\ }{\bf #1}}
\newcommand{\PL}[1]{{\it Phys.\ Lett.\ }{\bf #1}}
\newcommand{\CMP}[1]{{\it Commun.\ Math.\ Phys.\ }{\bf #1}}
\newcommand{\MPL}[1]{{\it Mod.\ Phys.\ Lett.\ }{\bf #1}}
\newcommand{\IJMP}[1]{{\it Int.\ J. Mod.\ Phys.\ }{\bf #1}}
\newcommand{\PR}[1]{{\it Phys.\ Rev.\ }{\bf #1}}
\newcommand{\PRL}[1]{{\it Phys.\ Rev.\ Lett.\ }{\bf #1}}
\newcommand{\PTP}[1]{{\it Prog.\ Theor.\ Phys.\ }{\bf #1}}
\newcommand{\PTPS}[1]{{\it Prog.\ Theor.\ Phys.\ Suppl.\ }{\bf #1}}
\newcommand{\AP}[1]{{\it Ann.\ Phys.\ }{\bf #1}}
\newcommand{\ATMP}[1]{{\it Adv.\ Theor.\ Math.\ Phys.\ }{\bf #1}}

\begin{thebibliography}{100}
%
\bibitem{MAL} J. Maldacena, 
        The large $N$ limit of superconformal field theories 
        and supergravity, \ATMP{2} (1998) 231, hep-th/9711200. 
\bibitem{GKP} S.S. Gubser, I.R. Klebanov and A.M. Polyakov, 
        Gauge theory correlators from noncritical string theory, 
        \PL{B428} (1998) 105, hep-th/9802109. 
\bibitem{WITTEN} E. Witten, 
        Anti de Sitter space and holography, 
        \ATMP{2} (1998) 253, hep-th/9802150. 
\bibitem{AGMOO} O. Aharony, S.S. Gubser, J. Maldacena, H. Ooguri and 
        Y. Oz, Large $N$ field theories, string theory and gravity, 
        {\it Phys.\ Rept.\ }{\bf 323} (2000) 183, hep-th/9905111. 
%
\bibitem{PS} J. Polchinski and M.J. Strassler, 
        The string dual of a confining four-dimensional 
        gauge theory, hep-th/0003136. 
\bibitem{MYERS} R.C. Myers, Dielectric-branes, 
        {\it JHEP }{\bf 12} (1999) 022, hep-th/9910053. 
\bibitem{TVR} W. Taylor and M. Van Raamsdonk,
        Multiple D$p$-branes in weak background fields, 
        \NP{B573} (2000) 703, hep-th/9910052. 
%
\bibitem{GP} M. Gra\~na and J. Polchinski,
        Supersymmetric three-form flux perturbations on AdS${}_5$, 
        \PR{D63} (2001) 026001, hep-th/0009211.
\bibitem{NT4} M. Nishimura and Y. Tanii, Three-form flux with 
        ${\cal N} = 2$ supersymmetry on AdS${}_5$ $\times$ S${}^5$, 
        {\it JHEP} {\bf 03} (2003) 019, hep-th/0212337. 
%
\bibitem{NISHI} M. Nishimura, 
        The string dual of an ${\cal N} = (4,0)$ two-dimensional gauge 
        theory, {\it JHEP} {\bf 07} (2001) 020, hep-th/0105170.
%
\bibitem{MS} J. Maldacena and A. Strominger, 
        AdS${}_3$ black holes and a stringy exclusion principle, 
        {\it JHEP} {\bf 12} (1998) 005, hep-th/9804085. 
\bibitem{DKSS} S. Deger, A. Kaya, E. Sezgin and P. Sundell, 
        Spectrum of $D = 6$, $N = 4b$ supergravity on AdS${}_3$ 
        $\times$ $S^3$, \NP{B536} (1998) 110, hep-th/9804166. 
\bibitem{BOER} J. de Boer, Six-dimensional supergravity on 
        ${\rm S}^3 \times {\rm AdS}_3$ and 2d conformal field theory, 
        \NP{B548} (1999) 139, hep-th/9806104. 
\bibitem{APT} G. Arutyunov, A. Pankiewicz and S. Theisen, 
        Cubic couplings in $D = 6$ ${\cal N} = 4b$ supergravity 
        on AdS${}_3$ $\times$ S${}^3$, 
        \PR{D63} (2001) 044024, hep-th/0007061.
%
\bibitem{CREMMER} E. Cremmer,
        Supergravities in 5 dimensions, 
        in {\it Superspace \& Supergravity}, eds.\ S.W. Hawking 
        and M. Ro\v cek (Cambridge Univ.\ Press, 1981). 
\bibitem{TANII} Y. Tanii, 
        $N=8$ supergravity in six dimensions, \PL{B145} (1984) 197. 
%
\bibitem{FMS} D. Friedan, E.J. Martinec and S.H. Shenker, 
        Conformal invariance, supersymmetry and string theory, 
        \NP{B271} (1986) 93.
%
\bibitem{BKL} V. Balasubramanian, P. Kraus and A.E. Lawrence,
        Bulk versus boundary dynamics in anti-de Sitter spacetime, 
        \PR{D59} (1999) 046003, hep-th/9805171.
\bibitem{BKLT} V. Balasubramanian, P. Kraus, A.E. Lawrence 
        and S.P. Trivedi,
        Holographic probes of anti-de Sitter space-times, 
        \PR{D59} (1999) 104021, hep-th/9808017.
%
\bibitem{VAFA} C. Vafa, 
        Gas of D-branes and Hagedorn density of BPS states, 
        \NP{B463} (1996) 415, hep-th/9511088.
\bibitem{SV} A. Strominger and C. Vafa,
        Microscopic origin of the Bekenstein-Hawking entropy, 
        \PL{B379} (1996) 99, hep-th/9601029.
\bibitem{DMW} J.R. David, G. Mandal and S.R. Wadia,
        Microscopic formulation of black holes in string theory, 
        {\it Phys.\ Rept.\ }{\bf 369} (2002) 549, hep-th/0203048.
\end{thebibliography}
\end{document}